\documentstyle [12pt] {article}
\title{CAN A NUCLEUS BE LARGER THAN AN ATOM (QUANTUM LAST SUPPER-POSITION)}

\author{Vladan Pankovi\'c, Darko Kapor\\
Department of Physics, Faculty of Sciences, 21000 Novi Sad,\\ Trg
Dositeja Obradovi\'ca 4. , Serbia, vladan.pankovic@df.uns.ac.rs}

\date {}
\begin{document}
\maketitle \vspace {0.5cm}
 \vspace {0.5cm}

\begin {abstract}
In this work we consider an extraordinary quantum mechanical
effect when, roughly speaking, the nucleus of an atom becomes
(linearly) larger than the whole atom. Precisely, we consider
Helium ion (in the ground state of the electron) moving
translationally with the speed much smaller than speed of the
electron rotation. This translation, effectively, changes neither
the total momentum, nor the de Broglie wave length of the
electron, nor the linear size of the atom corresponding to the
diameter of the electron orbit. But, this translation implies a
small nucleus momentum and nuclear de Broglie wavelength almost
hundred times larger than the electron de Broglie wavelength. In
the measurement of the nucleus wavelength using a diffraction
apparatus with a characteristic length constant proportional to
the proposed nucleus wavelength, according to standard quantum
mechanical formalism, the nucleus behaves practically certainly as
a wave. Then the unique, irreducible linear characteristic size
for such a nucleus is de Broglie wavelength. Such a measurement
effectively influences neither the electron dynamics nor linear
size of the atom. This implies that, in such measurement, the size
of the nucleus is in one dimension larger than the whole atom,
i.e. electron orbital. All this corresponds metaphorically to the
famous Leonardo fresco "Last Supper" where Jesus' words coming
from the nucleus, i.e. center of the composition, cause an
expanding "superposition" or dramatic wave-like movement of the
apostles.
\end {abstract}

\vspace {1.5cm}

As it is well-known, according to the usual, classical
mechanically simplified, interpretation of the atom structure
based on the remarkable Rutherford scattering experiments, an atom
is similar to a microscopic solar planetary system. In the center
of the atom there is, like to Sun, a small in respect to the whole
atom (with linear size proportional to $10^{-15}m$ ), but massive
(few and more thousand times much massive than electron) complex
particle, nucleus (consisting of the protons and neutrons). Around
the nucleus there is the peripherical shell consisting of the
rotating, like to the planets around Sun, simple particles,
electrons (with orbit radius proportional to $10^{-10}m$) small
and light in respect to the nucleus.

In more accurate, Bohr-Sommerfeld-de Broglie, quasi-classical or
naïve quantum atomic theory, electrons can be alternatively or
complementarily interpreted as the de Broglie waves so that any
electron orbit proposed by Bohr quantization postulate has a
natural number of corresponding de Broglie wavelengths. Simply
speaking, here peripherical electron waves (for small quantum
numbers) obtain the linear sizes (wavelengths) proportional to the
linear size of the atom. (More accurately, for quantum number n
orbit radius is larger n-times than wavelength. It implies that,
for $n \sim 1$, orbit radius and wavelength, are proportional.)
But within the Bohr-Sommerfeld-de Broglie, quasi-classical or
naïve quantum atomic theory, there are no additional predictions
on the (linear size of) atomic nucleus. In respect to the linear
size of the whole atom nucleus stands very small.

In this work we shall consider the theory of an extraordinary
quantum mechanical effect when, roughly speaking, the nucleus of
an atom becomes (linearly) larger than the whole atom. More
precisely, we shall consider a once-ionized Helium atom in the
ground state of the rotating electron. This ion, as the whole,
moves translationally with very small speed, many times smaller
than speed of the electron rotation. This translation,
effectively, changes neither the total momentum, nor the de
Broglie wave length of the electron, nor the linear size of the
atom corresponding to the diameter of the electron orbit. Also,
this translation implies a small nucleus momentum and nuclear de
Broglie wavelength which is close to a hundred times larger than
the electron de Broglie wavelength. Further, one can perform
measurement of the nucleus wavelength using a diffraction
apparatus with a characteristic length constant proportional to
the proposed nucleus wavelength. In such measurement, according to
standard quantum mechanical formalism [1], [2] and its Copenhagen
interpretation [3], [4], the nucleus behaves practically certainly
as a wave. Then the unique, irreducible characteristic linear size
for such a nucleus is de Broglie wavelength. Such a measurement
effectively influences neither the electron dynamics nor linear
size of the atom. This implies that, in such measurement, the size
of the nucleus is in one dimension larger than the whole atom,
i.e. electron orbital. This is not contrary but complementary to
measurement of the linear size of the nucleus, as a particle, by
means of the remarkable Rutherford experiment. All this
corresponds metaphorically and conceptually to the famous Leonardo
fresco "Last Supper" where Jesus' words (Logos) coming from the
nucleus, i.e. center of the composition, cause an expanding
"superposition" or dramatic wave-like movement of the apostles
around Jesus [5].

As it is well-known [2] a once-ionized Helium atom behaves
similarly to the neutral Hydrogen atom in the sense that it holds
only one electron in the electron shell. For this reason the once
ionized Helium atom can be exactly quantum dynamically separated
in two non-entangled sub-systems, center of mass and relative
particle. Center of mass represents a free propagating quantum
(sub)system, while relative particle represents the  quantum
(sub)system in the spherically symmetric Coulomb electrostatic
potential.

Since nucleus mass (neglecting mass excess) $M_{n}$ is here 7348
or $\sim 10^{4}$ times larger than electron mass $m_{e}$, center
of mass can be, in an excellent approximation, represented by
nucleus only. Linear size of the nucleus determined by remarkable
Rutherford scattering experiments equals approximately
$10^{-15}m$. Simultaneously, for the same reason, relative
particle can be, in an excellent approximation, represented by
electron only.

Suppose that electron is in the ground state. Then, in the
quasi-classical approximation, i.e. according to Bohr-Sommerfeld
atomic theory, electron propagates along first Bohr orbit with
radius $R_{e}= \frac {\hbar^{2}}{2m_{e}}\simeq 0.26 10^{-10}m$ or
diameter $D_{e}=2R{e}\simeq 0.52 10^{-10}m$. Speed of the electron
equals $v_{e}= \frac {1}{4 \pi \epsilon_{0}}\frac
{2e^{2}}{\hbar}\simeq 4.4 10^{6}m/s$  and corresponding de Broglie
wave length, proportional to the first Bohr orbit diameter,
$\lambda_{e} = \frac {h}{m_{e}v_{e}} = 2\pi R_{e}= D_{e}\pi \simeq
1.63 10^{-10}m$. In this way linear size of the once ionized
Helium atom, equivalent to the electron first Bohr orbit diameter,
is proportional to electron de Broglie wave length on this orbit.
    Suppose that the once ionized Helium atom as the whole propagates
translationally with a speed $V_{n}$ much smaller than $v_{e}$,
i.e.
\begin {equation}
    V_{n}=10^{-x}v_{e}\simeq 4.4 \cdot 10^{6-x}m/s
\end {equation}
where x represents a number greater than 1 and, for realistic
situations, smaller than 6. Such propagation, in an excellent
approximation, does not change the momentum, orbit diameter and de
Broglie wave length of the electron.

Propagating nucleus obtains the de Broglie wave length
\begin {equation}
   \lambda_{n} = \frac {h}{M_{n}V_{n}}\simeq \frac {h}{10^{4}m_{e}V_{n}}=
   \frac {h}{10^{4-x}m_{e}v_{e}}= 10^{-4+x}\lambda_{e}  .
\end {equation}
Obviously for
\begin {equation}
    6 \geq x \geq 5
\end {equation}
nuclear de Broglie wave length (1) is close to a hundred times
larger than electron de Broglie wave length, i.e.
\begin {equation}
   \lambda_{n} \sim 10^{2}\lambda_{e}\sim 10^{-8}m .
\end {equation}
Simultaneously, under the same condition (3), nucleus speed (1)
becomes close to 100m/s, i.e.
\begin {equation}
     V_{n}\sim 100 m/s
\end {equation}
representing a very small, atypical, but technically realizable at
least in the principle, speed for atomic particles.

Expression (4) holds very interesting implications. Namely,
suppose that there is some diffraction apparatus with
characteristic length constant proportional to nuclear de Broglie
wavelength (4). By interaction with the translationally
propagating, once ionized Helium atom such diffraction apparatus
will cause diffraction of the nuclear de Broglie wave only without
practically any influence at the electron de Broglie wave
(propagating through apparatus like geometrical ray).

In connection with the detector mentioned diffraction apparatus
can be consequently considered as a typical measuring apparatus
for measurement of the wave characteristics of some quantum
system.

In our case given diffraction measuring apparatus measures wave
characteristic or wavelength of the nucleus and, of course, yields
result comparable with (8). Moreover, according to usual,
Copenhagen interpretation [3], [4] of the standard quantum
mechanical formalism [1], [2], that is in the excellent agreement
with subtle experimental facts [6], [7], it must be stated the
following. Within given measurement of the wave characteristics
nucleus represents a wave exclusively. Then linear size of the
nucleus corresponds to characteristic, irreducible linear size of
the wave, i.e. to nuclear de Broglie wavelength (4). (This value,
of course, according to Heisenberg momentum-coordinate uncertainty
relation, corresponds numerically to the particle coordinate
uncertainty.)

On the other hand, in our case, such diffraction measuring
apparatus practically does not any influence at the electron,
precisely electron orbit, momentum and de Broglie wave length on
the orbit. It implies that such measurement does not change linear
size of the once ionized Helium atom proportional to electron de
Broglie wave length.

Finally, it can be consequently concluded the following. Within
discussed measurement linear size of the nucleus (representing
nuclear de Broglie wave length) becomes approximately hundred
times larger than linear size of the whole atom (proportional to
electron de Broglie wave length on the first Bohr orbit). (This is
not contrary but complementary to measurement of the linear size
of the nucleus, as a particle, by means of the remarkable
Rutherford experiment. Namely, Rutherford experiment represents
typical measurement of the particle characteristics of the quantum
system.) Obtained effect can be metaphorically called "last
supper-position" or quantum Leonardo effect since all this
corresponds metaphorically and conceptually to famous Leonardo
fresco "Last Supper". This painting demonstrates how Jesus' words
(Logos), coming from nucleus, i.e. center of the composition,
cause an expanding "superposition" or dramatic wave-like movement
of the apostles around Jesus.

In conclusion the following can be shortly repeated and pointed
out. In this work we consider the theory of an extraordinary
quantum mechanical effect when, roughly speaking, the nucleus of
an atom becomes (linearly) larger than the whole atom. More
precisely, we consider a once-ionized Helium atom in the ground
state of the rotating electron. This ion, as the whole, moves
translationally with very small speed, many times smaller than
speed of the electron rotation. This translation, effectively,
changes neither the total momentum, nor the de Broglie wave length
of the electron, nor the linear dimension of the atom
corresponding to the diameter of the electron orbit. Also, this
translation implies a small nucleus momentum and nuclear de
Broglie wavelength which is close to a hundred times larger than
the electron de Broglie wavelength. Further, one can perform
measurement of the nucleus wavelength using a diffraction
apparatus with a characteristic length constant proportional to
the proposed nucleus wavelength. In such measurement, according to
standard quantum mechanical formalism and its Copenhagen
interpretation, the nucleus behaves practically certainly as a
wave. Then the unique, irreducible linear characteristic size for
such a nucleus is de Broglie wavelength. Such a measurement
effectively influences neither the electron dynamics nor linear
size of the atom. This implies that, in such measurement, the size
of the nucleus is in one dimension larger than the whole atom,
i.e. electron orbital. This is not contrary but complementary to
measurement of the linear size of the nucleus, as a particle, by
means of the remarkable Rutherford experiment. All this
corresponds metaphorically and conceptually to the famous Leonardo
fresco "Last Supper" where Jesus' words (Logos) coming from the
nucleus, i.e. center of the composition, cause an expanding
"superposition" or dramatic wave-like movement of the apostles
around Jesus.

Authors are very grateful to Professors. Drs. Nenad Simonovi\'c,
Petar Gruji\'c and Tristan $H\ddot {u}$bsch for illuminating
discussion, support and help.

\section {References}

\begin{itemize}

\item [[1]] P. A. M. Dirac, {\it Principles of Quantum Mechanics} (Clarendon Press, Oxford, 1958)
\item [[2]] A. Messiah, {\it Quantum Mechanics} (North-Holand Publ. Co., Amsterdam, 1970)
\item [[3]] N. Bohr, Phys.Rev. {\bf 48} (1935) 696.
\item [[4]] N. Bohr, {\it Atomic Physics and Human Knowledge} (John Wiley, New York , 1958)
\item [[5]] K. Clark, {\it Leonardo da Vinci} (Penguin Books, London, 1939, 1993)
\item [[6]] A. Aspect, P. Grangier, G. Roger, Phys. Rev. Lett. {\bf 47} (1981) 460
\item [[7]] A. Aspect, J. Dalibard, G. Roger, Phys. Rev. Lett. {\bf 49} (1982) 1804

\end {itemize}

\end {document}